# Analysis of MIMO Systems used in planning a 4G-WiMAX Network in Ghana

E.T. Tchao[1], K. Diawuo[2], W.K. Ofosu[3], E. Affum[4]

Department of Electrical/Electronic Engineering, Kwame Nkrumah Univ. of Science and Tech, Kumasi, Ghana[1, 2, 4]
Department of Electrical Engineering Technology, Penn State Wilkes-Barre, USA[2].

*Abstract*—with the increasing demand for mobile data services, Broadband Wireless Access (BWA) is emerging as one of the fastest growing areas within mobile communications. Innovative wireless communication systems, such as WiMAX, are expected to offer highly reliable broadband radio access in order to meet the increasing demands of emerging high speed data and multimedia services. In Ghana, deployment of WiMAX technology has recently begun. Planning these high capacity networks in the presence of multiple interferences in order to achieve the aim of enabling users enjoy cheap and reliable internet services is a critical design issue. This paper has used a deterministic approach for simulating the Bit-Error-Rate (BER) of initial MIMO antenna configurations which were considered in deploying a high capacity 4G-WiMAX network in Ghana. The radiation pattern of the antenna used in the deploying the network has been simulated with Genex-Unet and NEC and results presented. An adaptive 4x4 MIMO antenna configuration with optimally suppressed sidelobes has been suggested for future network deployment since the adaptive 2x2 MIMO antenna configuration, which was used in the initial network deployment provides poor estimates for average BER performance as compared to 4x4 antenna configuration which seem less affected in the presence of multiple interferers.

## I. INTRODUCTION

Broadband internet is quite expensive in Ghana and in many Sub-Saharan African countries and as such, subscribers in many African countries with high data demand have no option than to rely on network operators who offer poor Digital Subscriber Lines (DSL) access and long customer connection times. Broadband internet subscribers were relieved and looked forward to enjoying lasting high data rate service at cheaper cost when deployment of WiMAX begun in Ghana.

It is a fact that planning a new network with a limited number of subscribers is not the real problem. The difficulty is to plan a network that allows future growth and expansion to meet the high demand for mobile data services. In 2nd generation systems, coverage planning was the most important and also the sufficient issue for operating the network [1].

Coverage prediction and capacity estimation were mostly well separable. In 3G and 4G networks, users operate on the same frequency carrier and as such the number of simultaneous connections directly influences the system capacity. Because of the need for 3G and 4G networks to continue offering higher data rate service as the network grows, planning these networks cannot be separated into coverage and capacity estimation as there is the need to increase capacity to serve multiple services like speech, internet and high data rate interactive services [2].

Even though 4G networks need to increase capacity in order to serve the increasing user demand, no breakthroughs in coding or modulation schemes are to be expected and additional spectrum resources are scarce [1]. This development has stimulated research into interference cancellation techniques [3], multicell processing [4], and cognitive radio [5] to improve the spectral efficiency of today's 4G wireless networks. However, none of these techniques is likely to carry the expected increase in mobile data traffic alone and further Base Station densification seems necessary. Base Station densification results in interferences since most of these 4G technology reuses frequency [6]. The resulting interferences as Base Station densification increase in wireless communication systems has motivated studies into Multi-Input Multi-Output (MIMO) systems with interferences in [7-12].

It implies that, in order to plan the deployed WiMAX network in Ghana to continue providing its subscribers higher data rate services as subscribers grow exponentially on the network, more Base Stations will have to be deployed. These Base Stations densification will in turn lead to high level of interference in the network. These expected high levels of interference will undoubtedly reduce network performance of the WiMAX network which will be disappointing to subscribers in Ghana who were eagerly looking forward to a long term solution to the expensive and unreliable broadband services in the country.

In order to help maintain the present high network performance, BER analysis of the current MIMO system in the presence of multiple interference sources is very important. BER performance analysis plays an important role in helping broadband wireless networks achieve maximum capacity while maintaining an acceptable grade of service and good speech quality [12].

This paper studies the potential sources of interference in the network and presents analysis into the BER performance of the MIMO antenna configuration which was used in the initial network deployment in the presence of multiple interferers. The analytical expressions are validated against Monte Carlo numerical methods and results presented accordingly.

## II. INITIAL WIMAX NETWORK

The network under study covers 55 sq km in the central parts of Accra and Tema. This network was designed to maximize coverage and as such Base Stations have been optimally placed to augment coverage problems. Initial planning met the main requirements of achieving the simulated coverage area and low Co channel and Adjacent channel interference levels as predicted for maintaining good quality of service[13].

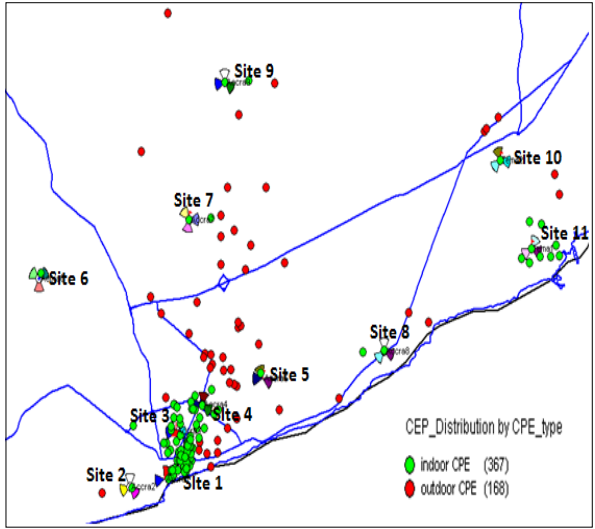

Fig.1. Distribution of CPE in the Deployed Network.

Currently, the number of Mobile Stations (MS) in the network has doubled and in order to serve the increasing demand by subscribers, more Base Stations will have to be deployed. Because of the short inter site distances, as seen in Table 1, deploying more Base Stations in the network will result in smaller cell sizes being created.

TABLE I. INTER-SITE DISTANCE FOR THE DEPLOYED NETWORK [14]

| Site A | Site B | Inter- site distances (km) |
|---|---|---|
| Site 1 | Site 2 | 1.938 |
| Site 1 | Site 3 | 1.072 |
| Site 1 | Site 4 | 2.423 |
| Site 4 | Site 5 | 1.563 |
| Site 5 | Site 6 | 6.20 |
| Site 5 | Site 8 | 7.542 |
| Site 5 | Site 7 | 8.590 |
| Site 7 | Site 9 | 5.342 |
| Site 6 | Site 7 | 5.734 |
| Site 10 | Site 11 | 4.494 |
| Site 10 | Site 8 | 23.33 |

Since a frequency reuse scheme of PUSC 1x3x3 was used in deploying the network in Figure 1 under a dense urban Sub-Saharan African environment, it is likely that as more Base Stations are added and cell sizes become smaller, there will an exponential increase in interferences and therefore the need for proper evaluation of the performance of the initial MIMO antenna configuration in the presence of multiple interferers.

## III. MIMO CHANNEL MODEL

One important practical implementation problem is that MIMO systems require knowledge of the channel conditions. Therefore, we consider a narrowband block fading channel with $n_R$ receiving antennas, $n_T$ transmit antennas from an interfering source, possibly representing several different users characterized by [15]:

$$y = \varsigma \omega + \varsigma_I \omega_I + z$$

Here, $y \in \mathbb{C}^{n_R}$ is the received signal vector. $\omega \in \mathbb{C}^{n_T}$ is the transmitted complex Gaussian distributed signal vector with zero mean and covariance $\beth = \mathrm{E}[\omega\omega^\varsigma], \omega_I \in \mathbb{C}^{n_I}$ is the interfering complex Gaussian distributed signal vector with zero mean and covariance $\beth_I = \mathrm{E}[\omega_I\omega_I^\varsigma]$, and $z \in \mathbb{C}^{n_R}$ is additive zero-mean white noise with entries $z_a \sim \mathcal{N}_c(0, 1)$. The channel matrices $\varsigma \in \mathbb{C}^{n_R \times n_T}$ and $\varsigma_I \in \mathbb{C}^{n_R \times n_I}$ has been model separately for correlated Rician fading. Thus, they can be written as

$$\begin{cases} \varsigma = \bar{\varsigma} + \wp^{1/2}\varsigma_w \Psi^{1/2} \\ \varsigma = \bar{\varsigma}_I + \wp_I^{1/2}\varsigma_{w,I}\Psi_I^{1/2} \end{cases} \quad (1)$$

Where $\bar{\varsigma}$ and $\bar{\varsigma}_I$ represent the mean values of $\varsigma$ and $\varsigma_I$, respectively, and are related to the presence of LOS components, $(\varsigma_w)_{ab}, (\varsigma_{w,I})_{ab} \sim \mathcal{N}_c(0, 1)$, and the positive semidefinite matrices $\Psi(\Psi_I)$ and $\wp(\wp_I)$ are the transmit signal (interference) and receive signal (interference) spatial correlation matrices respectively. The covariance between the different entries of $\varsigma$ and $\varsigma_I$ satisfies the identities

$$\begin{cases} \mathrm{cov}\big((\varsigma)_{ij}(\varsigma)_{i'j'}\big) = (\wp)_{ii'}(\Psi)^*_{jj'} \\ \mathrm{cov}\big((\varsigma_I)_{ij}(\varsigma_I)_{i'j'}\big) = (\wp_I)_{ii'}(\Psi_I)^*_{jj'}. \end{cases}$$

Extending these to multiple interfering transmitters, as in the case in network under discussion, (1) can be applied to model multiple interfering transmitter under the assumption of a common receive correlation matrix for each interfering source.

Indeed, assume that we have $N_I$ interfering users, so that the channel of the interferer is $i$ $(i = 1, \dots, N_I)$ of the form $\varsigma_I^i = \bar{\varsigma}^i + \wp^{1/2}W^i{\Psi_I^i}^{1/2}$ and

$y = \varsigma\omega + \sum_{i=1}^{N_I} \varsigma_I^i \omega_I^i + z.$ Then we set

$$\omega \triangleq \left[\omega_I^{1\mathrm{T}}, \dots, \omega_I^{N_I\mathrm{T}}\right]^\mathrm{T}, \quad \Psi \triangleq \bigoplus_{i=1}^{N_I} \Psi_I^i,$$

$W \triangleq [W^1, \dots, W^{N_I}]$, and $\bar{\varsigma} \triangleq [\bar{\varsigma}^1, \dots, \bar{\varsigma}^{N_I}].$

Different receive correlation matrices for the interfering transmitter can be modeled by introducing virtual delays in combination with a wideband channel model, as proposed in [15][16][17]. Following the Approach in [18] we define

$$\begin{cases} \check{\zeta} \triangleq \bar{\zeta} \beth^{1/2} \dot{\Psi} = \Psi^{1/2} \beth \Psi^{1/2} \\ \check{\zeta} \triangleq \check{\zeta}_I \beth_I^{1/2} \dot{\Psi}_I = \Psi_I^{1/2} \beth_I \Psi_I^{1/2} \end{cases} \quad (2)$$

Then, the transmitted signal and interference covariance matrices are implicitly accounted for into $\check{\zeta}, \check{\zeta}_I, \dot{\Psi}$ and $\dot{\Psi}_I$.

According to these definitions, the total received power is given by [18]

$$E[\|y\|^2] = \|\check{\zeta}\|^2 + \text{tr}(\wp)\text{tr}(\dot{\Psi}) + \|\check{\zeta}_I\|^2 + \text{tr}(\wp_I)\text{tr}(\dot{\Psi}_I) + n_R$$

Splitting the total received power components into direct and diffuse parts we obtain Rician factors

$$K = \frac{\|\check{\zeta}\|^2}{\text{tr}(\wp)\text{tr}(\Psi)} \quad K_I = \frac{\|\check{\zeta}_I\|^2}{\text{tr}(\wp_I)\text{tr}(\Psi_I)} \quad (3)$$

Then signal-to-noise and interference-to-noise power ratios can be defined as

$$\begin{cases} SNR = \frac{(K+1)\text{tr}(\wp)\text{tr}(\Psi)}{n_R} \\ INR = \frac{(K_I+1)\text{tr}(\wp)\text{tr}(\Psi_I)}{n_R} \end{cases} \quad (4)$$

It can be noticed that the definitions of the Rician factors and of the SNRs in (3) and (4) depend on the full transmit covariance matrices $\beth$ and $\beth_I$ and not only on their traces, unless they are proportional to the identity matrix. This will be an issue when the channel capacity of the MIMO channel in the presence of multiple interferers is evaluated against the SNR.

## IV. WiMAX ANTENNA FEATURES

An Adaptive Antenna Systems (AAS) was used in deploying the WiMAX network under discussion. The advantage in using adaptive antennas is that, it utilizes beam-forming techniques for focusing and directing the wireless signal between the base station and the receiver station. This reduces interference from other external sources and noises, as the beam is focused directly between two points.

The use of these techniques provides advantages on the basis of coverage, self-installation, power consumption, frequency re-use and bandwidth efficiency. Use of beam form techniques and MIMO under WiMAX reduces interference while increasing throughput and efficiency [19].

Antennas used in wireless communications have unwanted upper sidelobes. Sidelobes are known sources of interference in wireless networks. When the antenna has no downtilt, the upper sidelobes travel upwards and there is little chance of interference.

However, when downtilt is applied to the antenna as in the case of the deployed network, these unwanted upper sidelobes are now directed towards neighboring cells as shown in Figure 2. This has the potential of causing interference in the network since frequency is being reused.

Antenna pattern for the coverage simulations has been summarized in table 2 below. According to [20], all BS antenna elements have the beam pattern defined by 3GPP2. It is given by the formula.

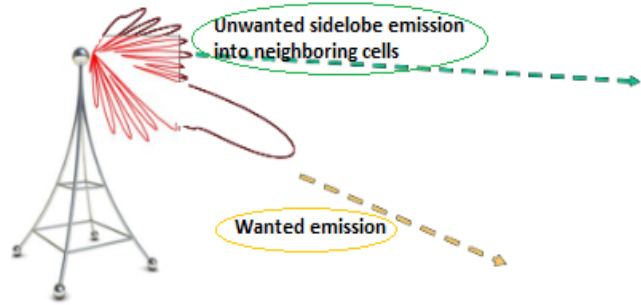

Fig.2. Antenna Radiating Pattern scenarios

$$G(\theta) = G_{max} + \max\left[-12\left(\frac{\theta}{\theta_{3dB}}\right)^2, -G_{FB}\right]$$

Gmax in (dBi) is the maximum antenna gain in the boresight direction. θ is the angle of arrival relative to boresight, $\theta_{3dB}$ is the 3dB beam width. $G_{FB}$ is the front-to-back ratio in dBi

In order to understand the effect multiple Base Station deployments will cause to the overall network performance of the MIMO systems, the radiating pattern of the antenna has been modeled in the next section and discussed

## V. SIMULATION RESULTS

Simulation of the radiation pattern of the WiMAX antenna has been done with Genex-Unet and NEC and shown in Figures 3, 4 and 5. The simulation parameters have been summarized in Table 2

From the non-optimal unsuppressed sidelobes as seen in the antenna's vertical radiation pattern in Figure 5, each BS in the network is a potential source of interference.

Performing a bit-error-rate simulation can be a lengthy process. In order to make sure that our results are statistically significant and allows for the different MIMO configurations to be compared in a fair manner, a Monte Carlo approach was used to model the BER performance of a BS in the presence of 10 interfering Base Stations. The result of the simulation has been shown in Figure 6.

TABLE II. RADIATING PATTERN SIMULATION PARAMETERS

| Frequency Range | 2300 MHz ~2700MHz |
|---|---|
| VSWR | ≤1.5 |
| Input Impedance | 50Ω |
| Gain | 18 dBi±0.5dBi |
| Polarization | ±45° |
| Horizontal Beamwidth (3dB) | 60±5° |
| Vertical Beamwidth (3dB) | 7°±0.5° |
| Electrical Downtilt | 2° |
| Isolation Between Ports | ≥30dB |
| Front to Back Ratio | ≥30dB±2 dB |
| Cross Polarization Ratio | ≥18dB |
| Upper Sidelobe Suppression | ≥18dB |
| Null-Fill | ≤18dB |
| Max, power | 250W |

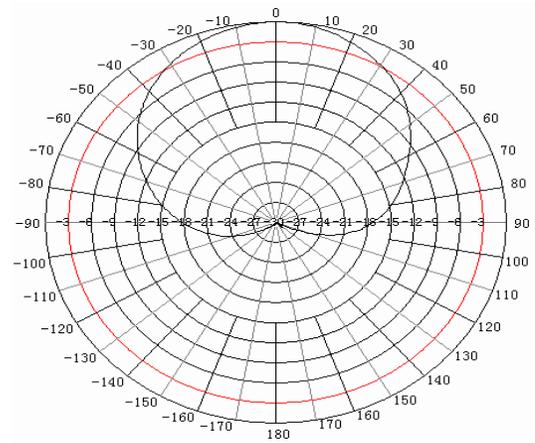

Fig.4. Horizontal Radiating Pattern

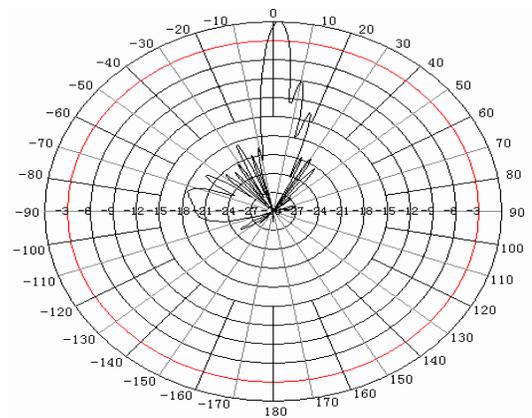

Fig.5. Vertical Radiating Pattern

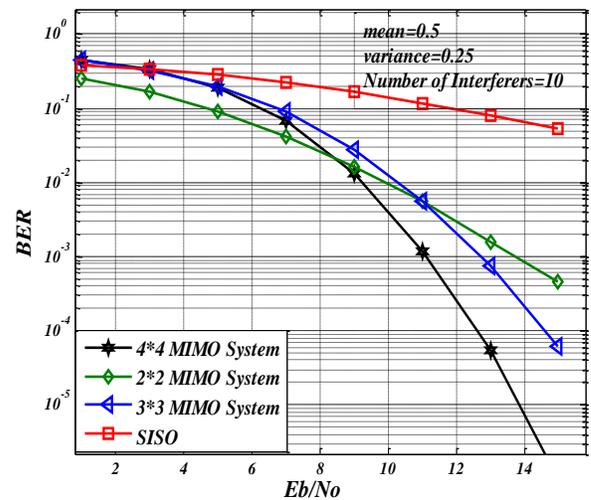

Fig.6. Performance of MIMO System in the presence of 10 interferers.

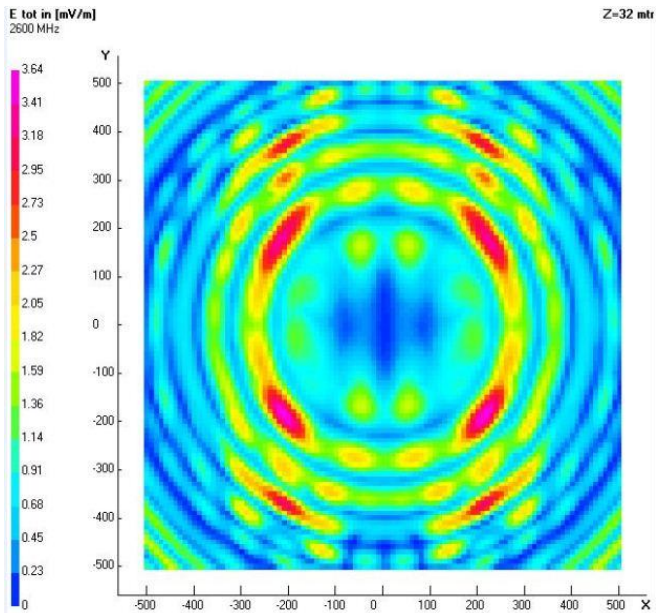

Fig.3. Antenna radiating pattern

A bounding technique of specifying the mean and variance was applied to ensure that the simulated BER estimates are fairly accurate.

## VI. Discussion Of Results

Interference reduces the integrity of signals travelling between transmitters and receivers. In cellular networks, this can often be caused by frequency reuse, therefore, stray radiation from one cell can disrupt transmission in other nearby cells using the same frequency. As the deployed WiMAX network uses a PUSC 1*3*3 frequency reuse scheme, the presence of unsuppressed sidelobes as seen from the antenna radiating pattern simulation in Figure 5 can significantly contribute to interference.

The BER simulation result using the Monte Carlo approach shows that the number of transmit ($n_T$) and receive ($n_R$) antennas affect the BER performance of the System. This complements the earlier results by Erwin *et all* on dealing with asymptotic statistics of the mutual information for correlated Racian Fading MIMO channels with interference. The results seems to be important for large number of antennas as the performance of the 4x4 MIMO system seemed the least affected by the variance and the number of interferers.

With the comparatively low performance of the 2x2 MIMO configurations, data transmitted on the network could be corrupted in the presence of multiple interferers. This may result in data having to be re-transmitted, and this reduces network performance. With Studies predicting a 1000X increase in data traffic from 2010-2020 [21], BER performance in the presence of interference is a critical design issue so as not to put severe strain on deployed 4G networks.

With the 4X4 MIMO configurations giving better performance in the presence of interference, it is recommended that future deployment of Base Stations in the network should use 4x4 MIMO antennas with optimized sidelobes suppression techniques.

## VII. Conclusion

Broadband wireless networks provide significant and promising growth marketplace for the telecom companies in Ghana to deliver a variety of applications and services to subscribers. In order for any network to sustain its initial high network performance after deployment, parameter analysis is necessary especially as the network grows and the number of potential interference to the network increases.

In the context of system performance, it can be concluded that 4x4 MIMO configuration provides satisfactory result than the initial 2x2 MIMO configuration in the presence of multiple interferers. A 4x4 MIMO antenna configuration with upper sidelobes optimally suppressed is recommended to be used in subsequent deployments especially in the presence of significant unsuppressed sidelobe emissions of the current antenna.

Evaluation of the effects of using antennas with upper sidelobes optimally suppressed against using the current antennas with typical upper sidelobe suppression technique will be presented when enough data is available from the field trial measurements.